\documentclass[10pt,prd,longbibliography,superscriptaddress]{revtex4-2}
\usepackage[T1]{fontenc}
\usepackage[english]{babel}
\usepackage{gensymb}

\usepackage{amsfonts,amsbsy,amssymb,amsmath,graphicx,float} 
\usepackage{color}

\graphicspath{{figures/}}

\begin{document}

\title{Bifurcation of dividing surfaces constructed from  a pitchfork  bifurcation of periodic orbits in a symmetric potential energy surface with a post-transition-state bifurcation}

\author{Matthaios Katsanikas}
\email{matthaios.katsanikas@bristol.ac.uk}
\affiliation{School of Mathematics, University of Bristol, \\ Fry Building, Woodland Road, Bristol, BS8 1UG, United Kingdom.}

\author{Makrina Agaoglou}
\email{makrina.agaoglou@icmat.es}
%\affiliation{School of Mathematics, University of Bristol, \\ Fry Building, Woodland Road, Bristol, BS8 1UG, United Kingdom.}
\affiliation{Instituto de Ciencias Matem{\'a}ticas, CSIC, C/Nicol{\'a}s Cabrera 15, Campus Cantoblanco,
28049 Madrid, Spain}

\author{Stephen Wiggins}
\email{s.wiggins@bristol.ac.uk}
\affiliation{School of Mathematics, University of Bristol, \\ Fry Building, Woodland Road, Bristol, BS8 1UG, United Kingdom.}

\begin{abstract}
In this work we analyze the bifurcation of dividing surfaces that occurs as a result of a pitchfork bifurcation of periodic orbits in a two degrees of freedom Hamiltonian System. The potential energy surface of the system that we consider has four critical points:two minima, a high energy saddle and a lower energy saddle separating two wells (minima). In this paper we study the structure, the range, and
the minimum and maximum extent of the periodic orbit dividing surfaces of the family of periodic orbits of the lower saddle as  a function of the total energy.
\end{abstract}

\maketitle


\section{Introduction}

The aim of this paper is to study a bifurcation
of  periodic orbit dividing surfaces that occurs as a result of a pitchfork bifurcation of periodic orbits. For this reason we  choose a pitchfork bifurcation of periodic orbits that takes place in a Hamiltonian system with two degrees of freedom. The potential energy surface (PES) that we use is symmetric about the $x$-axis and it  has  four critical points: two index-1 saddles and two minima. One index-1 saddle (the upper index-1 saddle) has higher energy than the other index-1 saddle (the lower index-1 saddle) and the lower index-1 saddle separates the two potential wells (the upper well and the lower well). Between the two index-1 saddles is a valley ridge inflection (VRI) point, where the shape of the potential near the VRI point changes from a valley to a ridge. This potential has been studied recently (\cite{collins2013nonstatistical,katsanikas2020c,Agaoglou2020,garcia2020tuning,agaoglou2021,douglas2021,crossley2021poincare}).

Dividing surfaces are surfaces of one less dimension than that of  the potential energy surface. These surfaces are important for the Transition State Theory (TST) (see \cite{Wigner38,waalkens2007}) in chemical reaction dynamics. In Hamiltonian systems with two degrees of freedom, the periodic orbit dividing surfaces are 2-dimensional surfaces in the energy surface. The construction of the periodic  orbit dividing surfaces is constructed through the classical method of \cite{Pechukas73,Pechukas77,Pechukas79,pollak1985,pechukas1981} in Hamiltonian systems with two degrees of freedom. The construction of the periodic orbit dividing surfaces in Hamiltonian with three or more degrees of freedom can be done through the method  of \cite{katsanikas2021ds,katsanikas2021bds}. 

An invariant, closed, oriented, co-dimension-2 submanifold of an energy level that can be spanned by two compact  codimension-1  submanifolds of unidirectional flux whose union, a dividing surface, separates locally the energy level into two components and it has no local  recrossing is called a transition state (see \cite{mackay2014bifurcations}).  The bifurcations of transition states (unstable periodic orbits and NHIMs) in chemical reaction dynamics have been  used in order to understand:  A) The breakdown of  transition state theory in collinear $HgI_2 \rightarrow HgI+I$ reaction \cite{burghardt1995molecular} and the hydrogen exchange reaction (see \cite{Inarrea11}). B) The influence of no-return transition states in the reaction rate  of   hydrogen exchange reaction \cite{Li09}. C) The loss and regain  of normal hyperbolicity of the transition states in a three-dimensional Hamiltonian model of the  hydrogen exchange reaction \cite{Allahem12}. D) The bridge between reactant and product channels in the collinear  $FH_2$ reactive system  \cite{founargiotakis1997bifurcation}. E) The study of systems  in which as the energy varies the transition state becomes  singular and then regains the normal hyperbolicity  with a change in differentiability as described through Morse bifurcations (see \cite{mackay2014bifurcations}). F) The separation of reactants and products and the flux through the dividing surface (see \cite{mauguiere2013bifurcations}). G) The transport of trajectories from the region of the higher index-1 saddle to the region of the top and bottom well in a symmetric potential energy surface with a post-transition-state bifurcation through a pitchfork bifurcation (see \cite{katsanikas2020c,Agaoglou2020}).

The basic and initial motivation of this work is to continue our previous work (\cite{katsanikas2020c,Agaoglou2020}) by investigating the effects of a pitchfork  bifurcation on the  structure  of the dividing surfaces of the top and bottom unstable periodic orbits (that bifurcate from the family of the lower saddle through a pitchfork bifurcation) that are responsible for the transport of the trajectories from the region of the higher index-1 saddle to the region of the top and bottom well (see \cite{katsanikas2020c,Agaoglou2020}). In all the papers that were mentioned in the previous paragraph except (\cite{mackay2014bifurcations},\cite{mauguiere2013bifurcations}), the studies were focused on  the applications of the bifurcations of periodic orbits  (see \cite{farantos2009energy} for a review of the many types of bifurcations in Hamiltonian systems with two and three degrees of freedom in chemical reaction dynamics) and NHIMs in  different settings of chemical reactions, but they did not focus on the effects of these bifurcations on the structure and geometry of dividing surfaces. In this paper, we deal with this problem and in particular with the evolution of the geometry of dividing surfaces as the energy increases and the family of periodic orbits of the lower saddle undergoes a pitchfork bifurcation in a symmetric potential energy surface with a post-transition-state bifurcation (\cite{katsanikas2020c,Agaoglou2020}). These surfaces are  spheres or ellipsoids before and after a bifurcation,  as is also shown for integrable cases of Hamiltonian systems \cite{mauguiere2013bifurcations}). They are presented as disks in the 3D projection $(x,y,p_x)$ of the phase space, before and after a pitchfork bifurcation in  a non-integrable Hamiltonian system with two degrees of freedom (\cite{lyu2021hamiltonian}). In our case, we will study in a more systematic way the change of the geometry of the dividing surfaces, before and after a pitchfork bifurcation,   investigating the  structure of the dividing surfaces in all 3D projections of the phase space of a non-integrable Hamiltonian system with two degrees of freedom. We will also describe  in detail  the structure of the dividing surfaces in a system in which the structure of dividing surfaces is more complicated (the dividing surfaces  are initially not disks  in the 3D projection $(x,y,p_x)$  but filamentary structures) than that of the dividing surfaces  that are found in the Hamiltonian model of \cite{lyu2021hamiltonian}. Furthermore,  we will follow the complex distortion of these surfaces that are initially ellipsoids or  filamentary structures in the 3D projections of the phase space as we change the energy. We emphasize the fact that  we investigated the distortion of the dividing surfaces (that are initially ellipsoids or filamentary structures) through a pitchfork bifurcation of periodic orbits, and not situations where the dividing surfaces change topology and become tori or other topological structures though a Morse bifurcation (see \cite{mackay2014bifurcations}). The study of topological changes of this kind (through a Morse bifurcation) is an interesting topic but out of the scope of this paper. We study, also the evolution of other geometrical characteristics,  such as the range and the maximum and minimum value of the dividing surfaces.

We describe the PES of our system in section \ref{model}. We describe the pitchfork bifurcation of periodic orbits of the lower saddle in section \ref{periodic}. Then in section \ref{results}, we describe our results about the structure of the diving surfaces. Finally, we discuss our results and we present our conclusions in the last section \ref{conclusions}.

\section{Model}
\label{model}

The analytical form of the PES that was inspired by \cite{collins2013nonstatistical} is the following:

\begin{equation}
V(x,y) = \frac{8x^3}{3} - 4x^2 + \frac{y^2}{2} + xy^2(y^2-2).
\label{PES}
\end{equation}

\begin{figure}[htbp] 
	\begin{center}
		\includegraphics[scale=2.2]{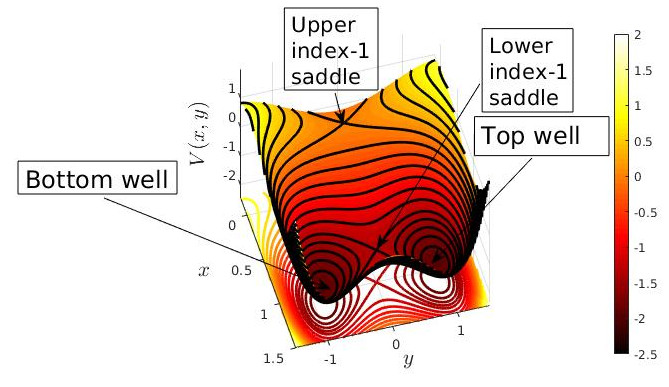} 
	\caption{Plot of the PES given in Eq. (\ref{PES}).}
    \end{center}
    \label{Pess}
\end{figure}

The PES has an upper index-1 saddle (exit/entrance channel) and a lower index-1 saddle, which is an energy barrier separating two potential wells. Moreover, the PES is symmetric about the x axis. A three dimensional plot of the PES is given in Fig. \ref{Pess}. The two dimensional (2D) Hamiltonian is given by

\begin{equation}\label{hamil}
  H(x,y,p_x,p_y) = \frac{p_x^2}{2} + \frac{p_y^2}{2} + V(x,y),   
\end{equation}

\noindent
and the corresponding Hamiltonian vector field (i.e. Hamilton's equations of motion) is:

\begin{equation}\label{ham1}
 \begin{cases}
   \dot{x} = p_x,\\
   \dot{y} = p_y,\\
   \dot{p}_x = 8x(1-x) + y^2(2-y^2), \\
   \dot{p}_y = y \left(4x(1-y^2)-1 \right).
  \end{cases}
\end{equation}

\noindent
Hamilton's equations conserve the total energy, which we will denote as $E$ throughout the text.

\section{Periodic Orbits}
\label{periodic}

\begin{figure}
 \centering
A)\includegraphics[scale=0.33]{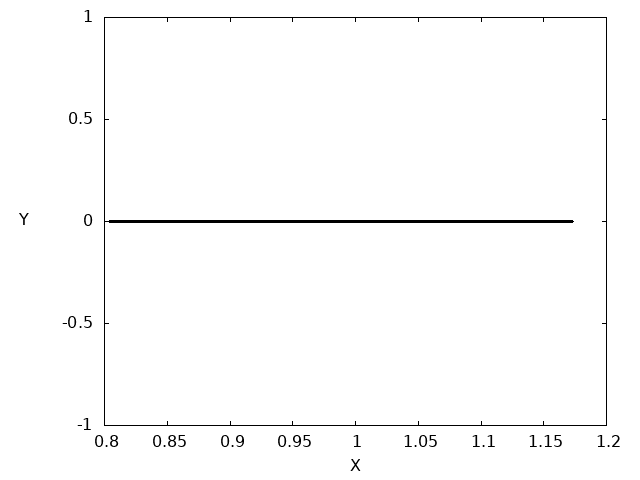}
B)\includegraphics[scale=0.33]{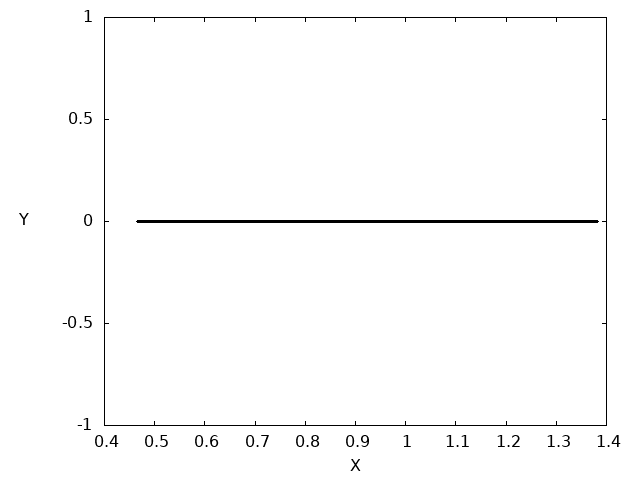}\\
C)\includegraphics[scale=0.33]{periodorbsadm12.png}
D)\includegraphics[scale=0.33]{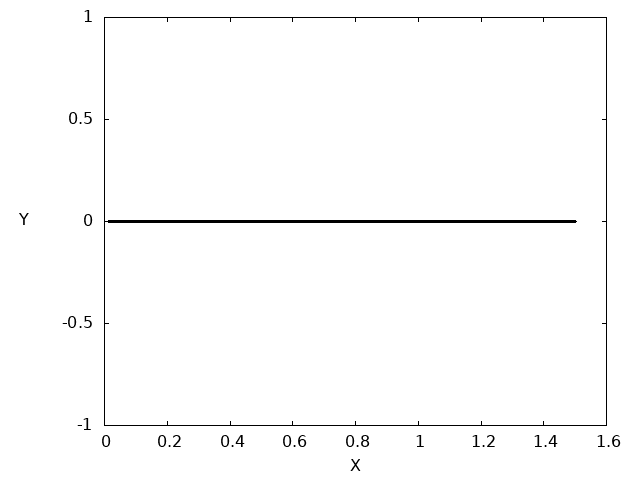}\\
\caption{2D projections of the periodic orbits of the central family in the configuration space for energies A) $-1.2$, B) $-0.6$, C) $-0.2$ and D) $-0.00055$}
\label{periodicorbitscentralxy}
\end{figure}

In this section, we  present a pitchfork bifurcation of the family of the lower saddle. As we explained in the previous section, we have two index-1 saddles, one for higher values of energy and one for lower values of energy. According to the Lyapunov subcenter theorem 
we have a family of periodic orbits associated with every index-1 saddle (see \cite{rabinowitz1982periodic,weinstein1973normal,moser1976periodic}). In this section we follow  the family of periodic orbits associated with the lower index-1 saddle, increasing the energy above the energy of the lower index-1 saddle. The periodic orbits  of this family lie on the $x$-axis (see Fig. \ref{periodicorbitscentralxy}).

\begin{figure}
 \centering
\includegraphics[scale=0.53]{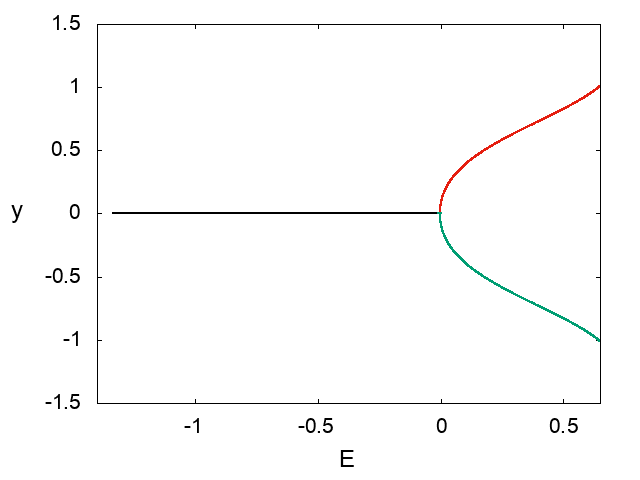}
\\
\caption{The evolution of the coordinate $y$ of periodic orbits on the Poincar\'e section $x=0.05$ versus energy (characteristic diagram). The family of periodic orbits of the lower saddle is depicted by black color and its bifurcating families, the families of the top and bottom unstable periodic orbits, are depicted by red and green color, respectively.  }
\label{char}
\end{figure}

\begin{figure}
 \centering
A)\includegraphics[scale=0.33]{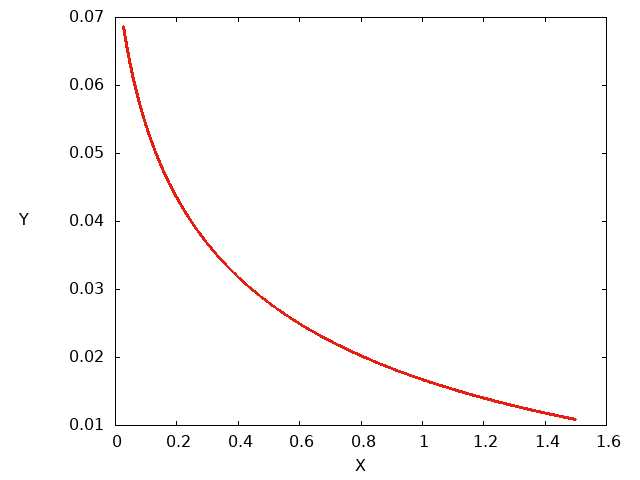}
B)\includegraphics[scale=0.33]{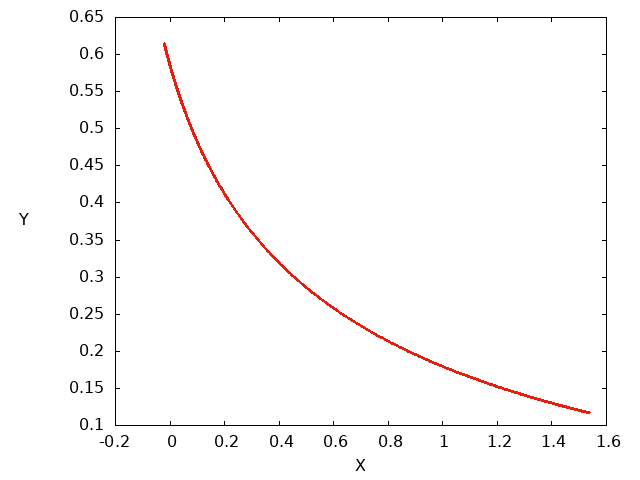}\\
C)\includegraphics[scale=0.33]{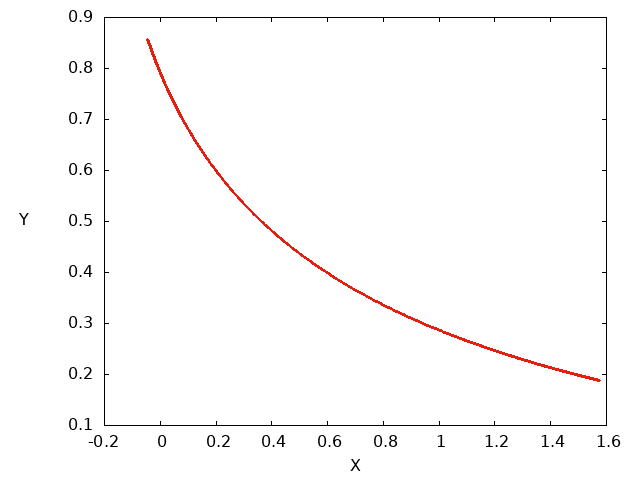}
D)\includegraphics[scale=0.33]{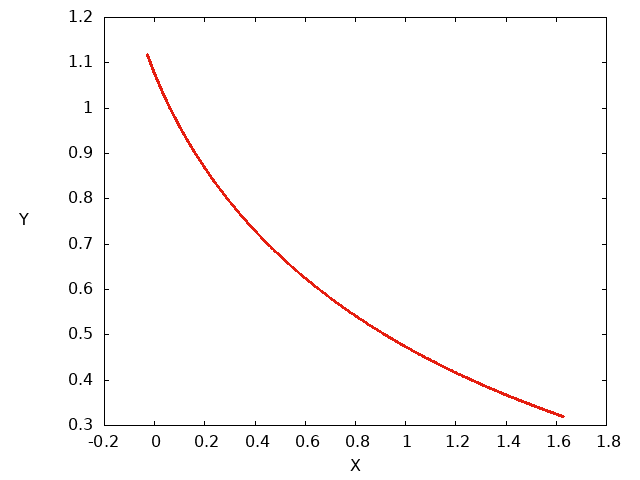}\\
\caption{2D projections of the periodic orbits of the top family in the configuration space for energies A) $-0.00055$, B) $0.2$, C) $0.4$ and D) $0.65$}
\label{periodicorbitstopxy}
\end{figure}

\begin{figure}
 \centering
A)\includegraphics[scale=0.33]{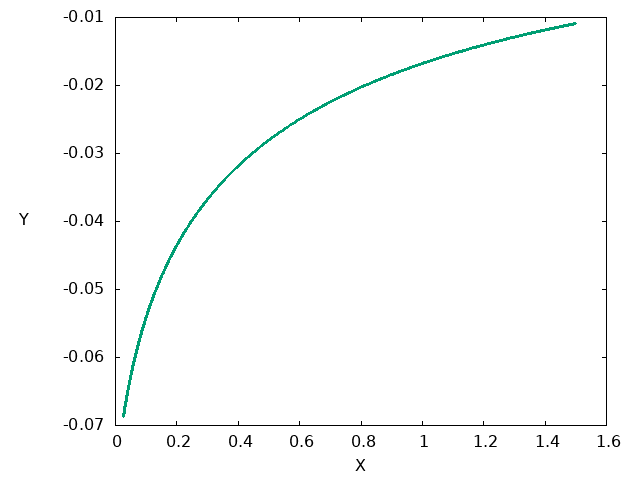}
B)\includegraphics[scale=0.33]{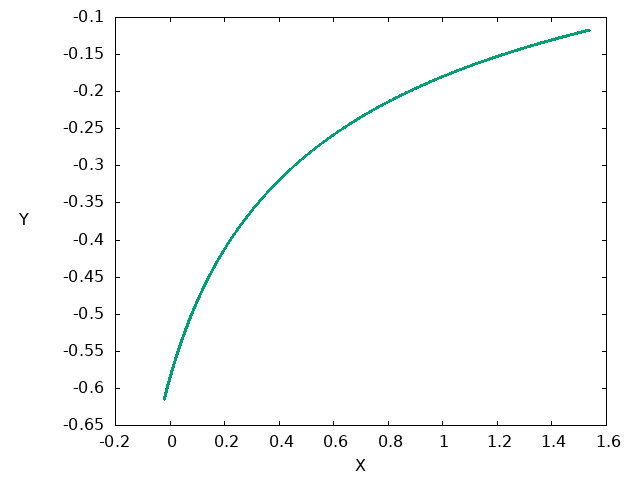}\\
C)\includegraphics[scale=0.33]{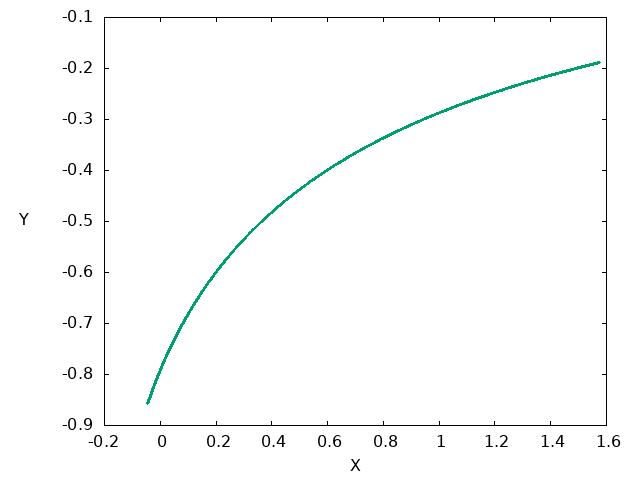}
D)\includegraphics[scale=0.33]{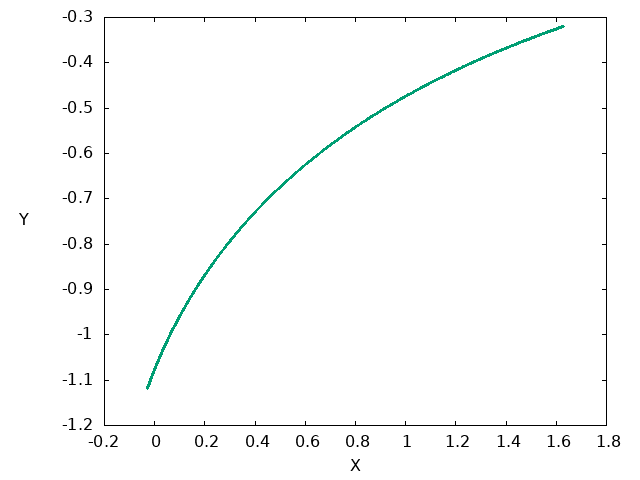}\\
\caption{2D projections of the periodic orbits of the bottom family in the configuration space for energies A) $-0.00055$, B) $0.2$, C) $0.4$ and D) $0.65$}
\label{periodicorbitsbottomxy}
\end{figure}

In the characteristic diagram (a diagram that shows the evolution of a coordinate of periodic orbits of a family in a Poincar\'e  section versus the energy of the system - see Fig. \ref{char}) we see that the family of periodic orbits associated  with the lower saddle undergoes a pitchfork bifurcation at $E=-0.00056$. At this value the family of  periodic orbits  associated with the lower index-1 saddles, that was initially unstable, becomes stable. At this point we have two new unstable families of periodic orbits which we refer to as the top and bottom  families of periodic orbits (see more details for the role of the these periodic orbits in the dynamics of the system in \cite{katsanikas2020c,Agaoglou2020,douglas2021,agaoglou2021}). We observe,  in Figs. \ref{periodicorbitstopxy}, \ref{periodicorbitsbottomxy}, that by increasing the energy the periodic orbits of these families become more elongated in the direction of the $y$-axis. In these figures we see that the POs of the top and bottom families are converging to the line $y=0$ from above and below, respectively.

\section{Results}
\label{results}
\begin{figure}
 \centering
A)\includegraphics[scale=0.53]{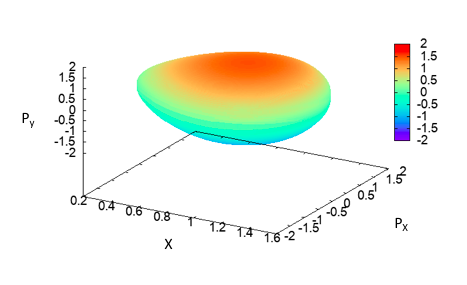}
B)\includegraphics[scale=0.53]{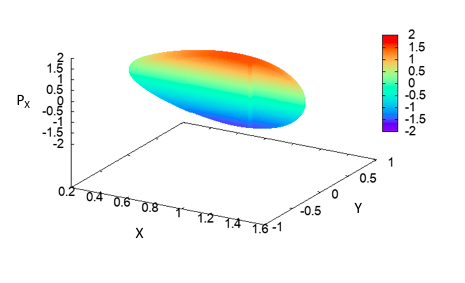}\\
\caption{3D projections $(x,p_x,p_y)$ (panel A) and $(x,y,p_x)$ (panel B) of the periodic orbit dividing surface of the family of the lower saddle for $E=-0.2$. }
\label{3d-02}
\end{figure}

\begin{figure}
 \centering
\includegraphics[scale=0.53]{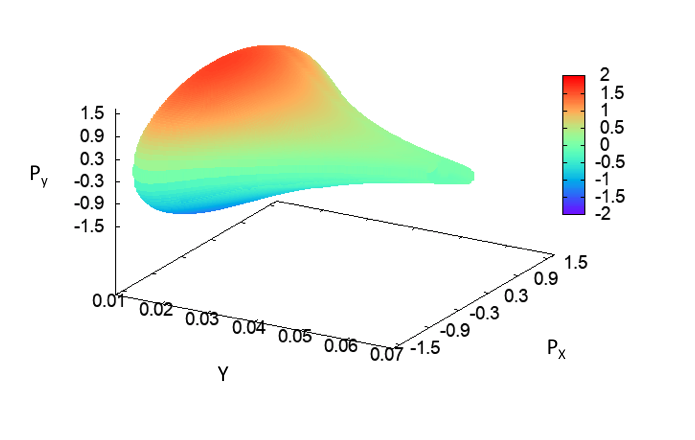}
\caption{3D projection $(x,p_x,p_y)$  of the dividing surface that is constructed by the family of the top unstable periodic orbits for $E=-0.00055$. }
\label{3d-00055}
\end{figure}

\begin{figure}
 \centering
A)\includegraphics[scale=0.43]{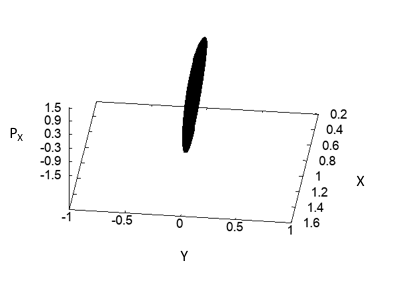}
B)\includegraphics[scale=0.43]{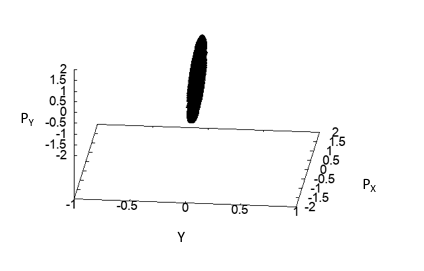}\\
C)\includegraphics[scale=0.43]{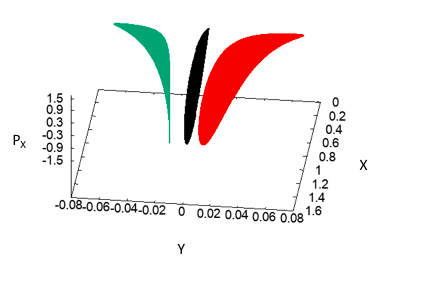}
D)\includegraphics[scale=0.43]{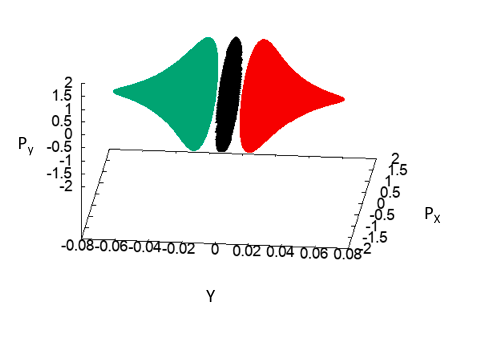}\\
E)\includegraphics[scale=0.38]{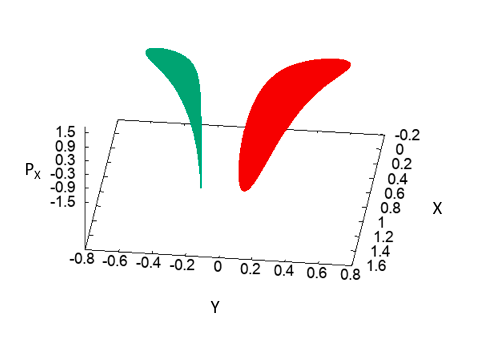}
F)\includegraphics[scale=0.38]{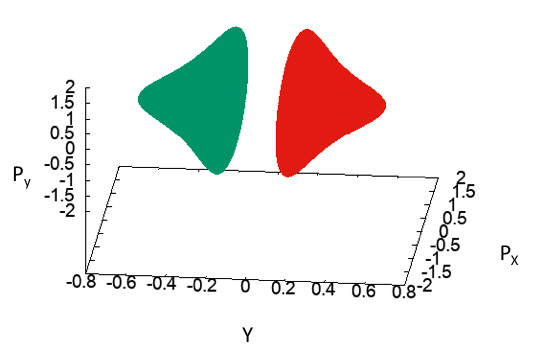}\\
\caption{3D projections $(x,y,p_x)$ (left column of the panels) and $(y,p_x,p_y)$ (right column of the panels) of the periodic orbit dividing surface of the family of the lower saddle (with black color) and of the dividing surfaces that are constructed  from the periodic orbits of the top (with red color) and bottom (with green color) unstable periodic orbits
for $E=-0.2$ (first row), $E=-0.00055$ (second row) and $E=0.2$ (third row). We encountered the same structures, as these in the 3D subspace  $(x,y,p_x)$, in the 3D subspace  $(x,y,p_y)$.  }
\label{3d-all}
\end{figure}

\begin{figure}
 \centering
A)\includegraphics[scale=0.33]{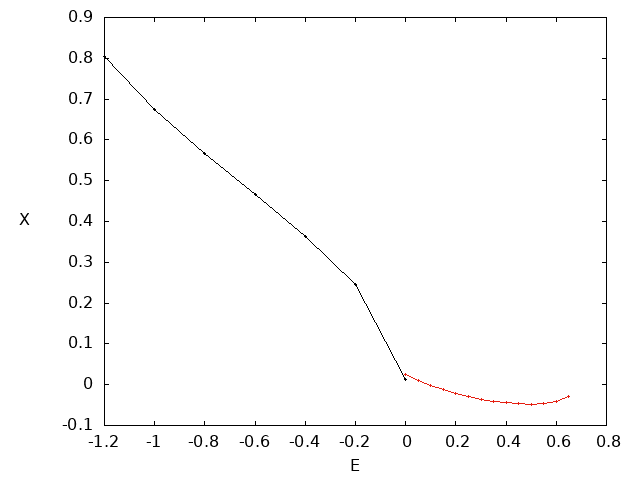}
B)\includegraphics[scale=0.33]{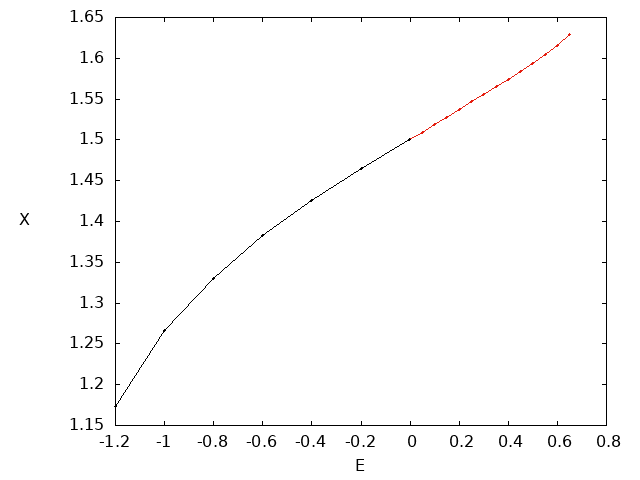}\\
C)\includegraphics[scale=0.33]{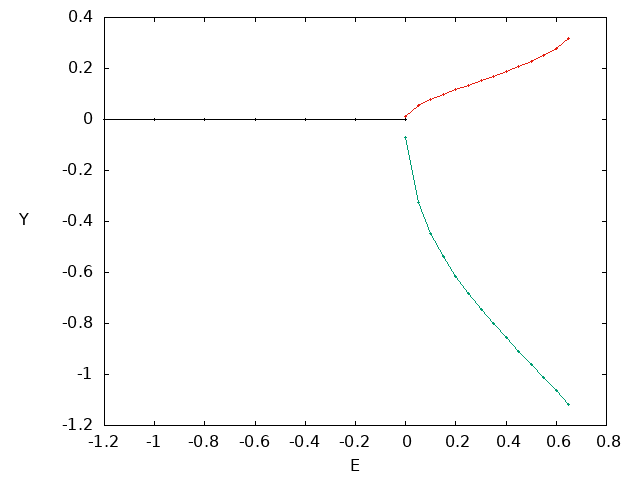}
D)\includegraphics[scale=0.33]{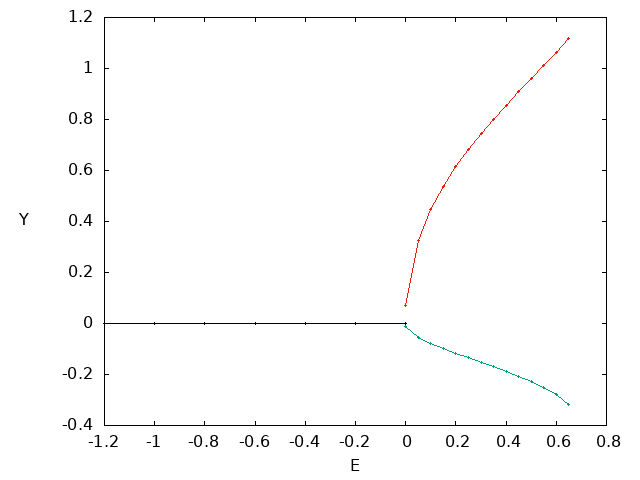}\\
E)\includegraphics[scale=0.33]{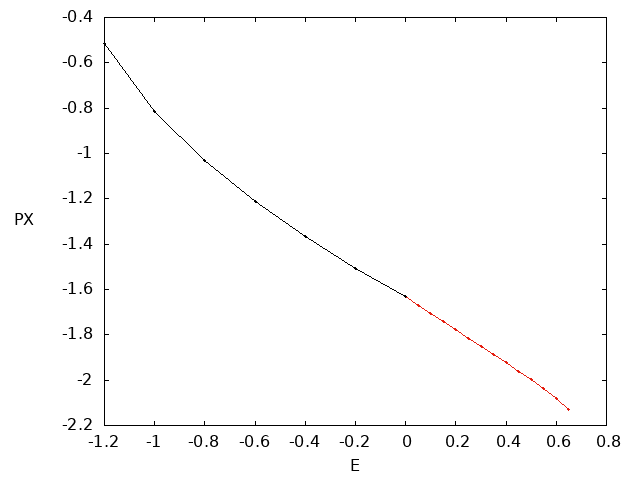}
F)\includegraphics[scale=0.33]{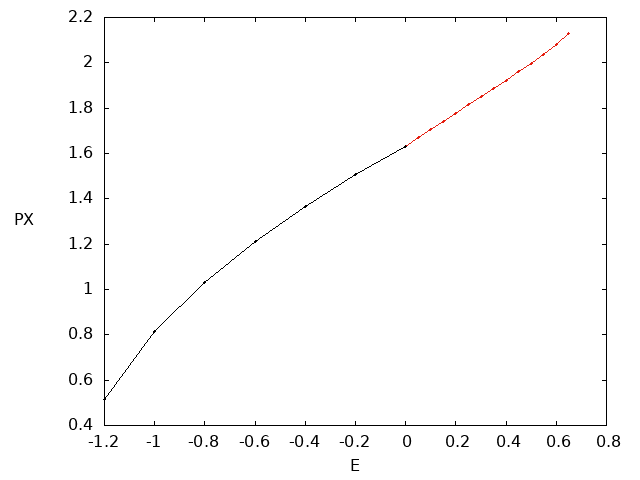}\\
\caption{Diagrams for the A) mininum and B) maximum of the  X coordinate of the central  (black), top (red) and bottom (green) families versus the energy, C) mininum and D) maximum of the  y coordinate of the central, top and bottom families versus the energy and E) mininum and F) maximum of the  PX coordinate of the  central, top and bottom families versus the energy.}
\label{minmax}
\end{figure}

\begin{figure}
 \centering
A)\includegraphics[scale=0.5]{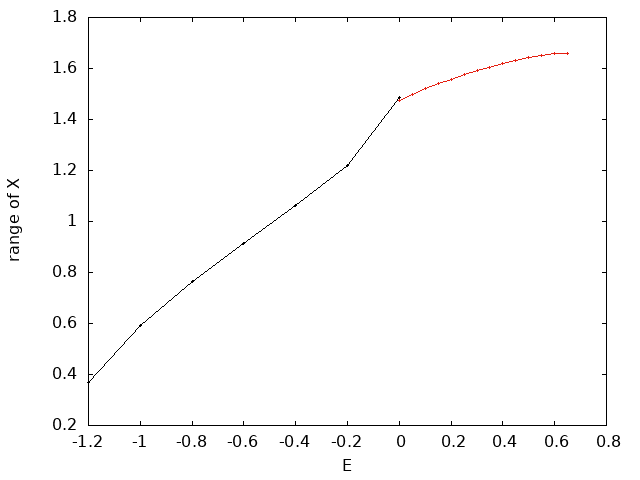}\\
B)\includegraphics[scale=0.5]{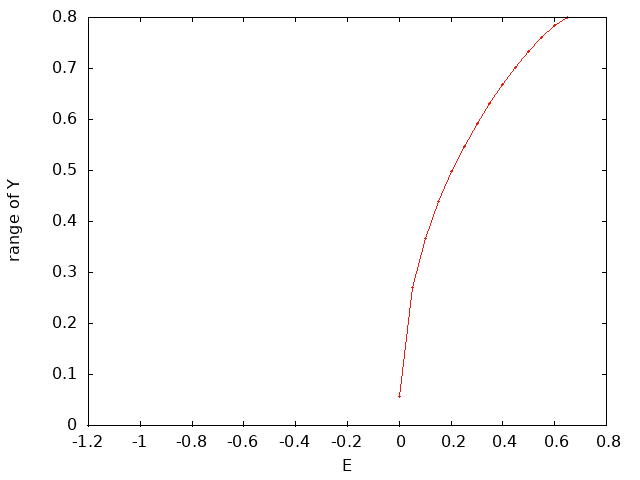}\\
C)\includegraphics[scale=0.5]{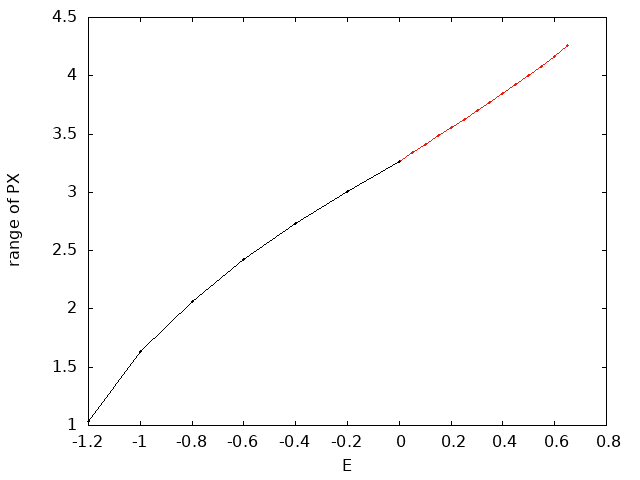}
\caption{A) Range of the X coordinate of the central family (black line), top family (red line) and bottom family (green line) versus the energy, B) Range of the Y coordinate of the central family (black line), top family (red line) and bottom family (green line) versus the energy and C) Range of the PX coordinate of the central family (black line), top family (red line) and bottom family (green line) versus the energy}
\label{range}
\end{figure}

\begin{figure}
 \centering
A)\includegraphics[scale=0.33]{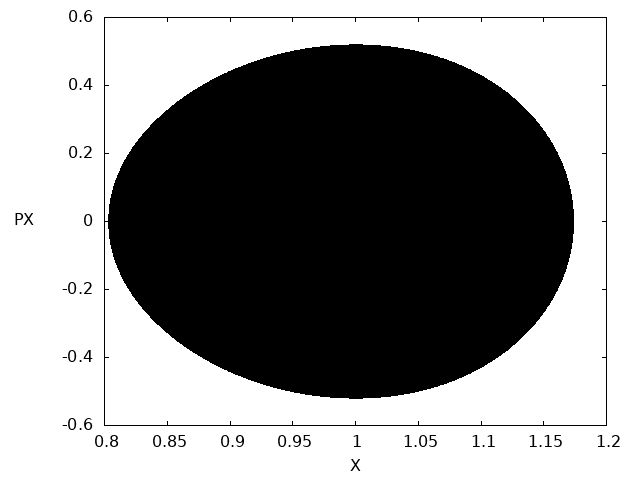}
B)\includegraphics[scale=0.33]{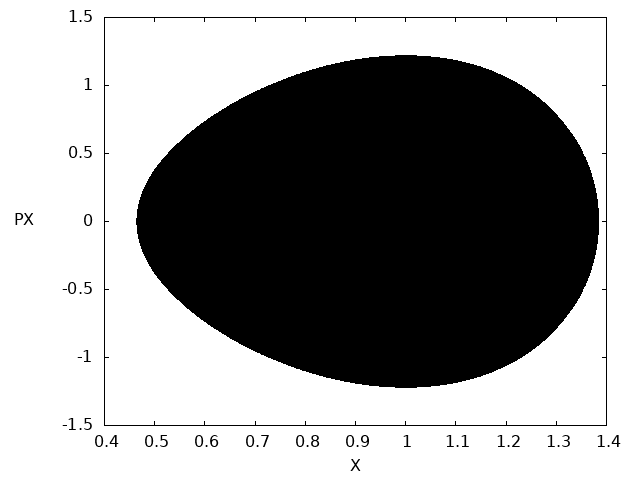}\\
C)\includegraphics[scale=0.33]{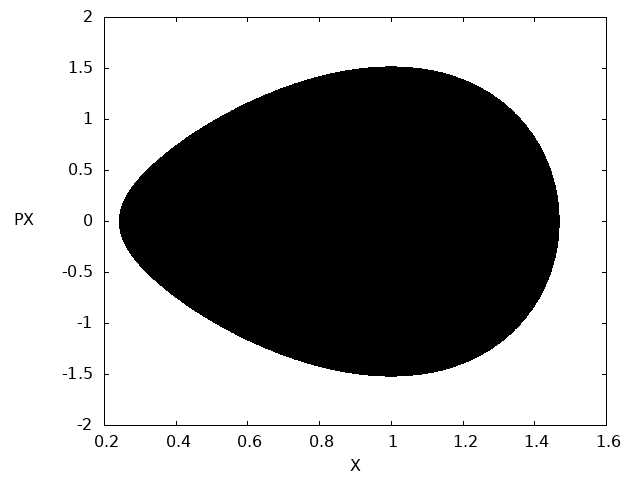}
D)\includegraphics[scale=0.33]{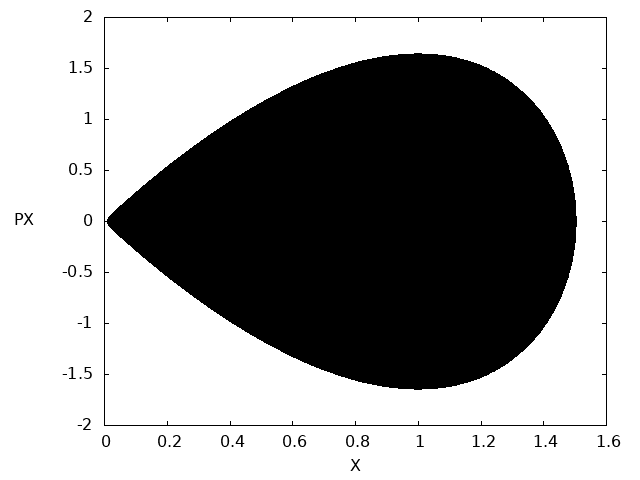}\\
\caption{2D projections of the dividing surfaces of the central family in the X-PX space for energies A) $-1.2$, B) $-0.6$, C) $-0.2$ and D) $-0.00055$}
\label{divcentralxpx}
\end{figure}

\begin{figure}
 \centering
A)\includegraphics[scale=0.33]{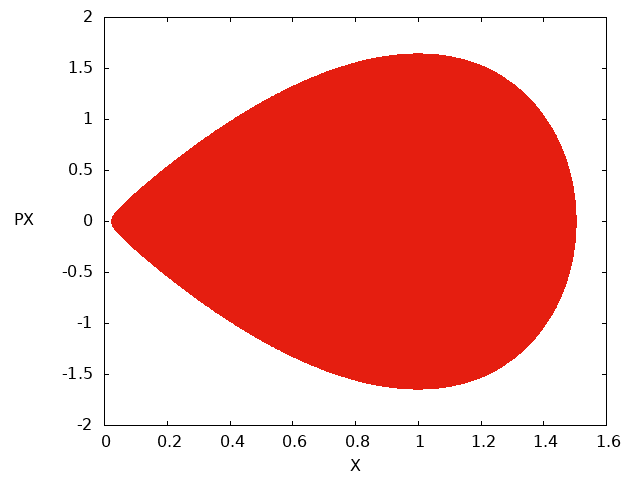}
B)\includegraphics[scale=0.33]{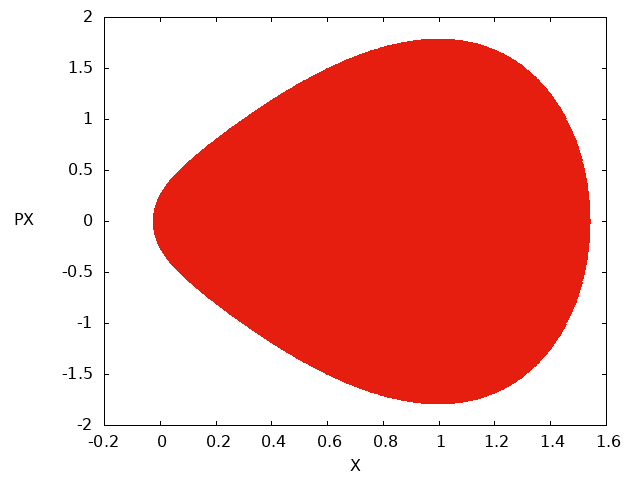}\\
C)\includegraphics[scale=0.33]{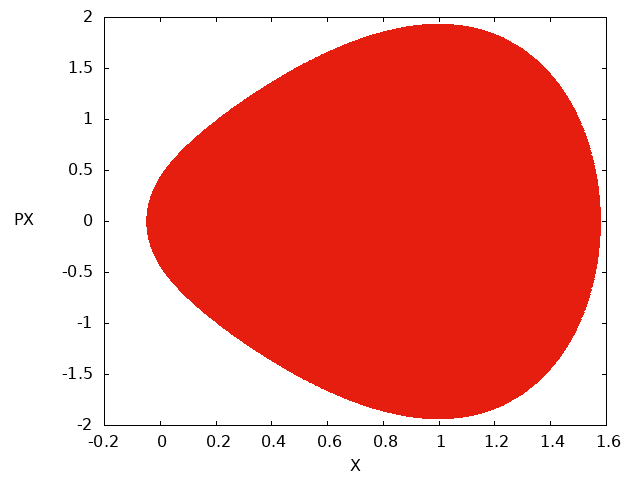}
D)\includegraphics[scale=0.33]{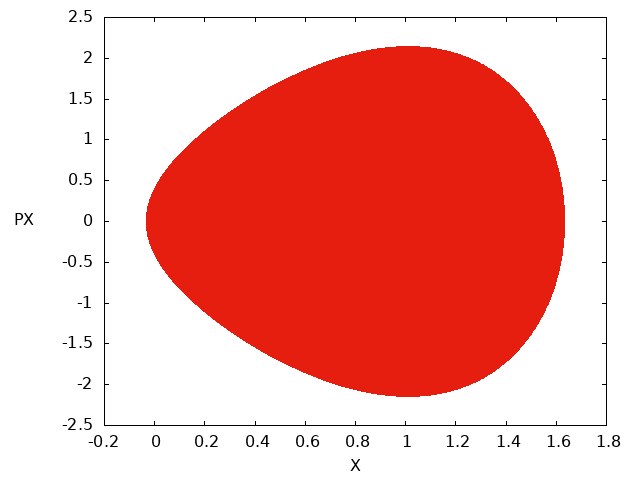}\\
\caption{2D projections of the dividing surfaces of the top family in the X-PX space for energies A) $-0.00055$, B) $0.2$, C) $0.4$ and D) $0.65$}
\label{divtopxpx}
\end{figure}

\begin{figure}
 \centering
A)\includegraphics[scale=0.33]{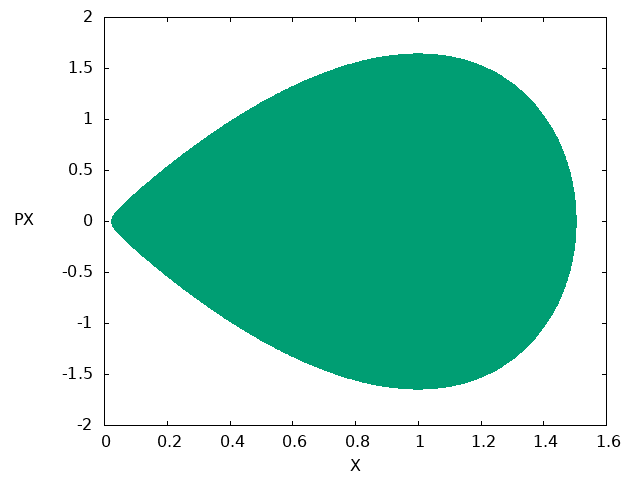}
B)\includegraphics[scale=0.33]{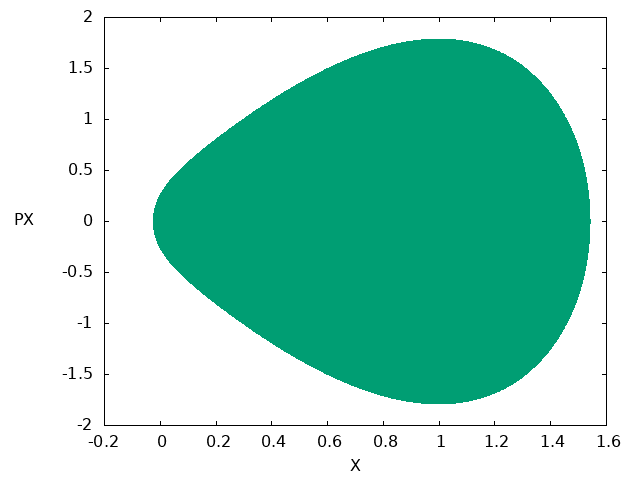}\\
C)\includegraphics[scale=0.33]{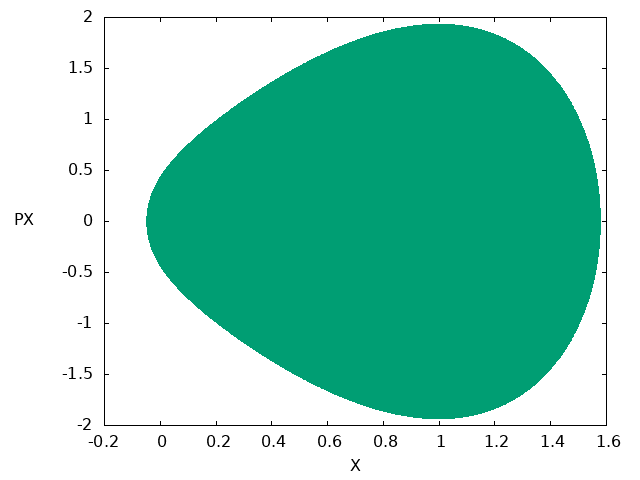}
D)\includegraphics[scale=0.33]{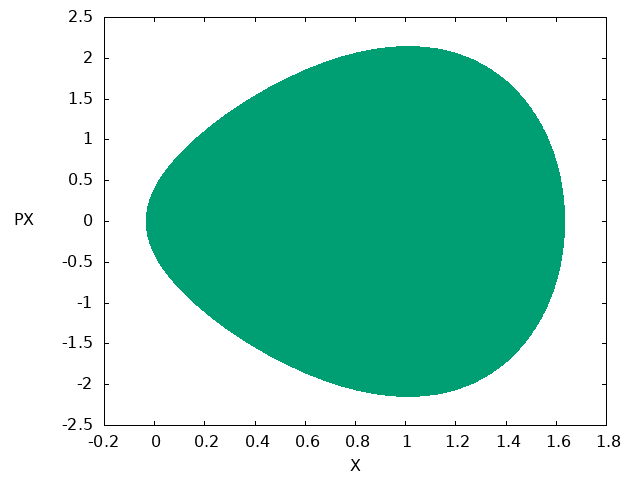}\\
\caption{2D projections of the dividing surfaces of the bottom family in the X-PX space for energies A) $-0.00055$, B) $0.2$, C) $0.4$ and D) $0.65$}
\label{divbottomxpx}
\end{figure}

\begin{figure}
 \centering
A)\includegraphics[scale=0.33]{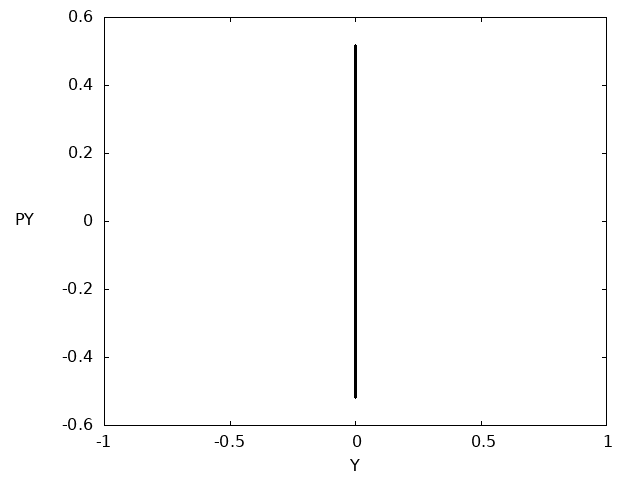}
B)\includegraphics[scale=0.33]{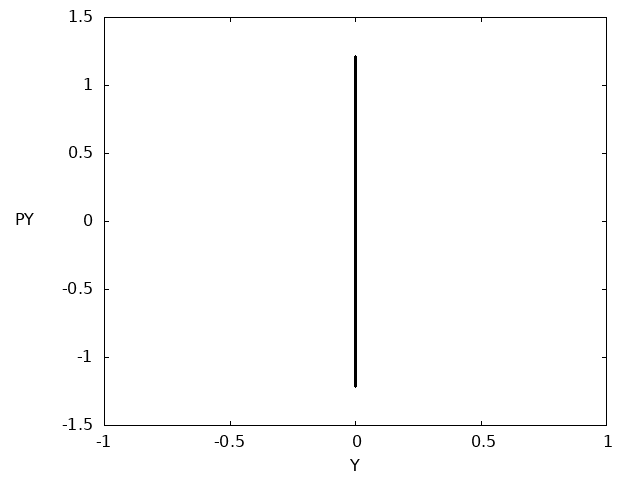}\\
C)\includegraphics[scale=0.33]{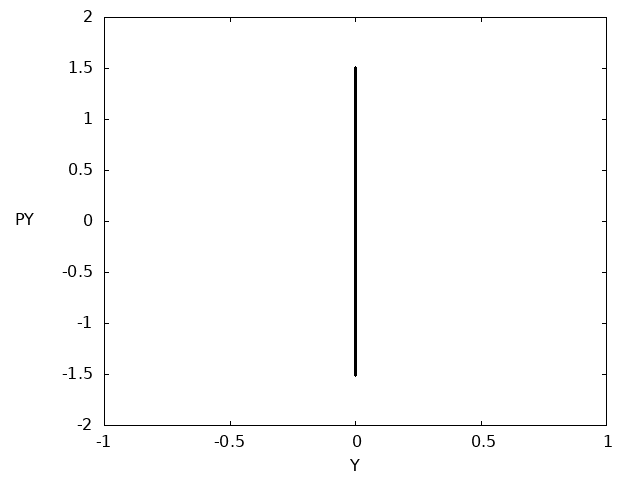}
D)\includegraphics[scale=0.33]{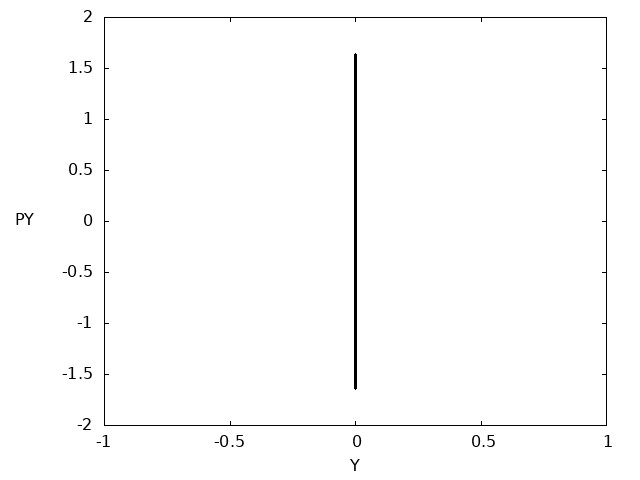}\\
\caption{2D projections of the dividing surfaces of the central family in the Y-PY space for energies A) $-1.2$, B) $-0.6$, C) $-0.2$ and D) $-0.00055$}
\label{divcentralypy}
\end{figure}

\begin{figure}
 \centering
A)\includegraphics[scale=0.33]{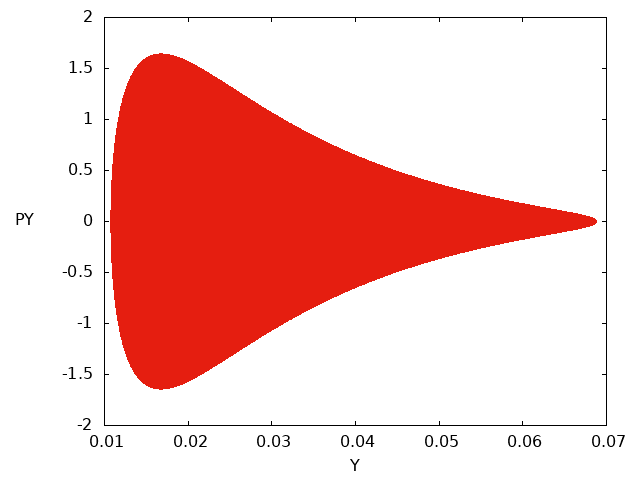}
B)\includegraphics[scale=0.33]{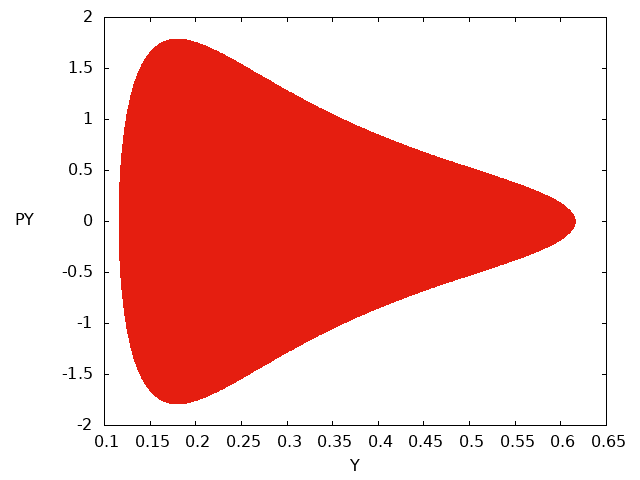}\\
C)\includegraphics[scale=0.33]{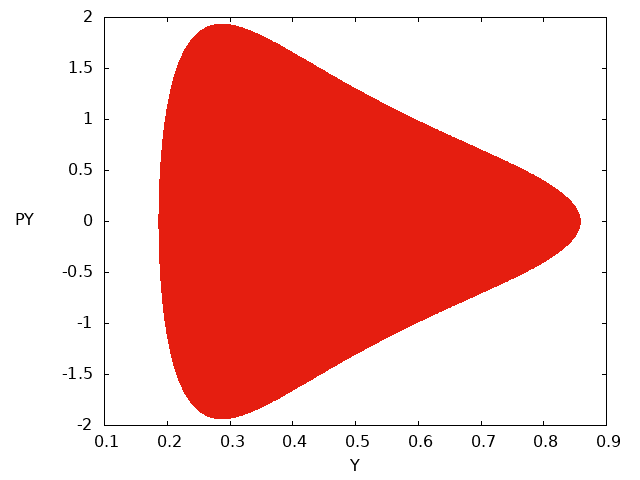}
D)\includegraphics[scale=0.33]{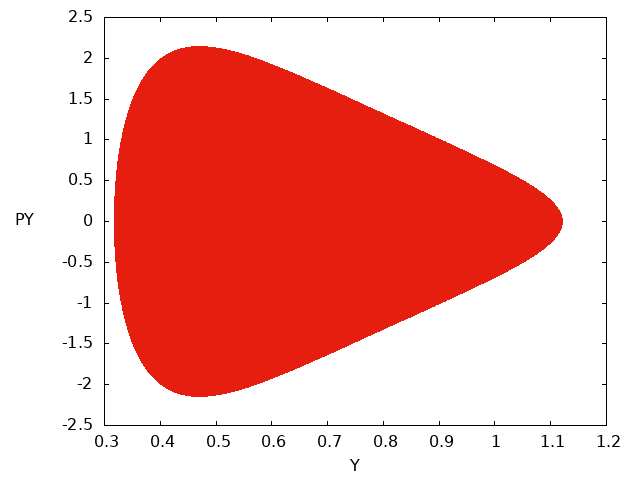}\\
\caption{2D projections of the dividing surfaces of the top family in the Y-PY space for energies A) $-0.00055$, B) $0.2$, C) $0.4$ and D) $0.65$}
\label{divtopypy}
\end{figure}

\begin{figure}
 \centering
A)\includegraphics[scale=0.33]{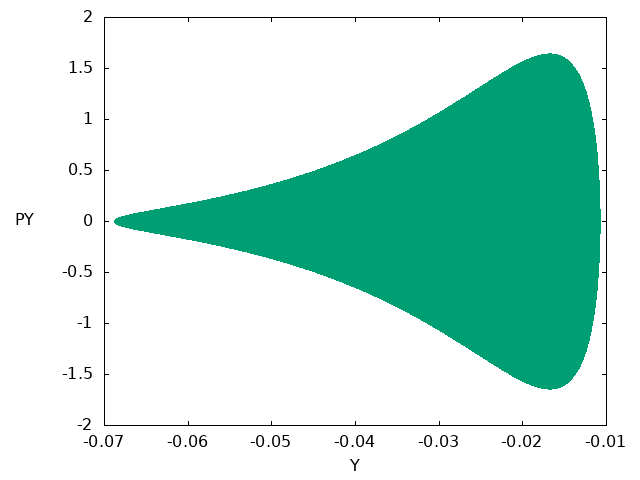}
B)\includegraphics[scale=0.33]{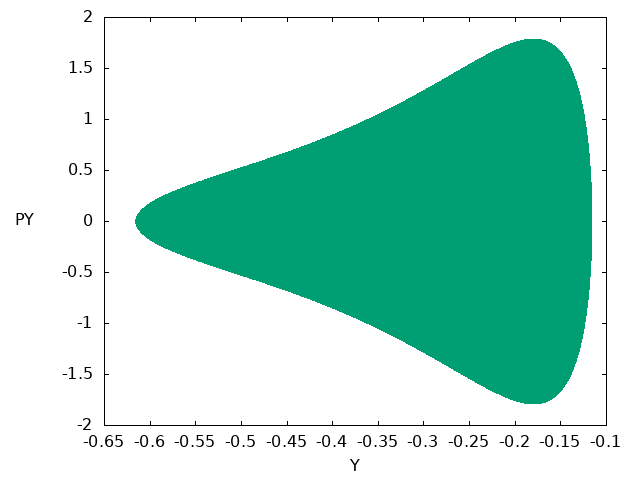}\\
C)\includegraphics[scale=0.33]{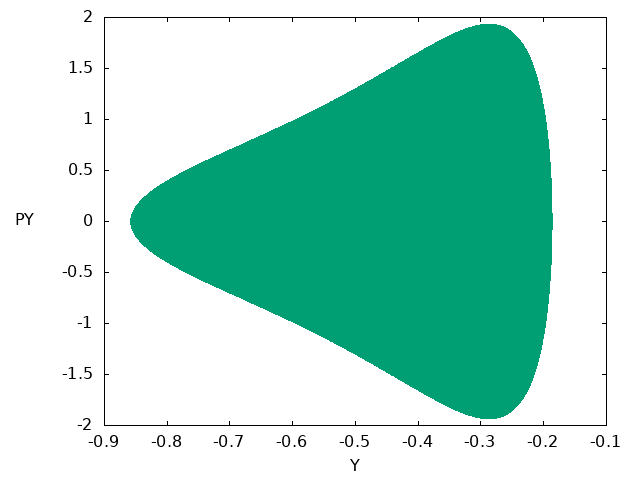}
D)\includegraphics[scale=0.33]{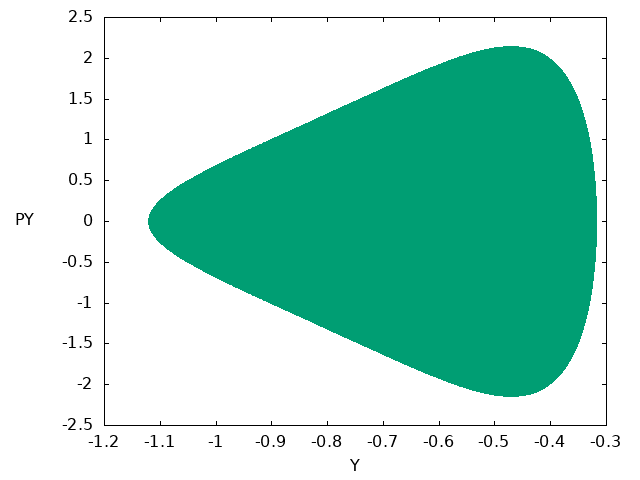}\\
\caption{2D projections of the dividing surfaces of the bottom family in the Y-PY space for energies A) $-0.00055$, B) $0.2$, C) $0.4$ and D) $0.65$}
\label{divbottomypy}
\end{figure}

Initially (for values of energy above the energy of the lower saddle, $E=-4/3$ and before the pitchfork bifurcation for $E=-0.00056$)  we have the family of  periodic orbits of the lower saddle that gives rise to a family  of dividing surfaces. For the construction of the periodic orbit dividing surfaces we use he classical algorithm of \cite{Pechukas73,Pechukas77,pechukas1981,Pollak78,pollak1985}, that is presented in our previous papers (see  (\cite{douglas2021,ezra2018sampling}). After the pitchfork bifurcation of the periodic orbits of the family of the lower saddles (see the previous section) 
we have two new bifurcating families, the top and bottom unstable periodic orbits. These families give rise to two new families of dividing surfaces. In this section, we study the evolution of the structure and the range and of the minimum  and maximum of the dividing surfaces that are constructed from the periodic orbits of the lower saddle and its bifurcations as energy is varied:

\begin{enumerate}

\item {\bf The structure of the dividing surfaces:} The periodic orbit dividing surfaces that are constructed from the periodic orbits of the lower saddles are presented as ellipsoids in the $(x,p_x,p_y)$  projection (see Fig. \ref{3d-02}).
This ellipsoid becomes more stretched and sharp as we increase the energy (see the 2D projections $(x,p_x)$ of the dividing surfaces in Fig. \ref{divcentralxpx}). Furthermore, these dividing surfaces are presented as  filamentary structures, that lie on the plane $y=0$, in the $(x,y,p_x)$ and  $(x,y,p_y)$ subspaces of the phase space (see panel B of Fig. \ref{3d-02} and panel A of Fig. \ref{3d-all}) and in the  $(y,p_x,p_y)$  subspace of the phase space (see panel B of Fig. \ref{3d-all}). This filamentary structure of these surfaces is because of the morphology of the periodic orbits of the lower saddles in the configuration space (it lies on the line $y=0$ - see the previous section). This is the reason that the 2D projections $(y,p_y)$ of the dividing surfaces of the  family of lower saddle (Fig. \ref{divcentralypy}) is the line $y=0$ for all energies.

As the energy is increased above that of the pitchfork bifurcation, in addition to the dividing surfaces of the family of the lower saddles,  we have two other families  of dividing surfaces. One family corresponds to the dividing surfaces  of the top unstable periodic orbits and the other to the dividing surfaces of the bottom unstable periodic orbits. The dividing surfaces of these families are on both sides of the dividing surfaces of the family corresponding to the lower saddles in the 3D projections  $(x,y,p_x)$, $(x,y,p_y)$ and $(y,p_x,p_y)$ (see panels C and D of Fig. \ref{3d-all}). The dividing surfaces of the bifurcating families in the 3D projections, $(x,y,p_x)$ and $(x,y,p_y)$, are presented as filamentary structures  that are curved in the direction of positive $y$-values (the dividing surfaces of the top unstable periodic orbits)  or on the direction of negative  $y$-values (the dividing surfaces of the bottom unstable periodic orbits). The structure of these dividing surfaces in the 3D subspace  
$(y,p_x,p_y)$  is similar to that of an ellipsoid that is a little sharper in one direction (see for example Fig. \ref{3d-00055}). In addition, the dividing surfaces of the family of the lower saddle coincide with the dividing surfaces of the two bifurcating families in the $(x,p_x,p_y)$ subspace of the phase space and they have the same structure (an ellipsoid) as  that of panel A of  Fig. \ref{3d-02}.

For values of energy larger than  0 (the energy of the upper saddle) the family of periodic orbits of the lower saddle vanishes. Consequently, the family of dividing surface  of the lower saddle vanishes too. This means that the dividing surfaces of the family of the lower saddle do not exist between the dividing surfaces of the bifurcating families  in  the 3D projections  $(x,y,p_x)$, $(x,y,p_y)$ and $(y,p_x,p_y)$ (see the panels E and F of Fig. \ref{3d-all}). We notice also that the distance between the dividing surfaces of the top and bottom unstable periodic orbits,  in  the 3D projections  $(x,y,p_x)$, $(x,y,p_y)$ and $(y,p_x,p_y)$, increases as the energy increases (compare the panels C and D of Fig. \ref{3d-all} with the panels E and F of Fig. \ref{3d-all}).

As we increase the energy, the dividing surfaces of the bifurcating families in the 3D subspace  $(x,p_x,p_y)$  become  more round/ellipsoidal as the energy increases (see the 2D projections $(x,p_x)$ of the dividing surfaces of the top and bottom unstable periodic orbits in Figs. \ref{divtopxpx},     \ref{divbottomxpx}). Furthermore, we observe  a decrease of the sharpness of the ellipsoidal dividing surfaces of the bifurcating families in the 3D subspace  $(y,p_x,p_y)$, as we increase the energy  (see  the 2D projections $(y,p_y)$ of the dividing surfaces of the top and bottom unstable periodic orbits in Figs. \ref{divtopypy}, \ref{divbottomypy}).

\item {\bf Maximum and Minimum of the dividing surfaces:} In Fig. \ref{minmax} we present in the left column the minimum of the dividing surfaces with respect to the  $x$ coordinate, $y$ coordinate and $p_x$ coordinate (panels A, C and E respectively) of the  family of the lower saddle (black line), top family (red line) and bottom family (green line), respectively. Similarly, in the right column of the figure we show the maximum of the dividing surfaces with respect to the $x$ coordinate, $y$ coordinate and $p_x$ coordinate (panels B, D and F respectively) of the family of the lower saddle (black line), top family (red line) and bottom family (green line), respectively. Note that we did not show the minimum and maximum of $p_y$ coordinate because it is determined from the other three coordinates through the Hamiltonian of the system. We observe that in the panel A the minimum of the  family of the lower saddle  is decreasing as the energy increases, as does does the minimum of the top and bottom families so that they coincide. In the panels B, E and F we notice that there is a continuation of the minima and maxima of the family of the lower saddle to the other two bifurcating families (they coincide with each other). Finally in the panel C we show that the minimum of the family of the lower saddle is always zero whilst the minimum of the top family is increasing and the minimum of the bottom family is decreasing. The same happens in the panel D for the maximum.

\item {\bf The range of the dividing surfaces:} In Figure \ref{range} we illustrate the range of the $x$ coordinate, $y$ coordinate and $p_x$ coordinate (panels A, B and C, respectively) of the dividing surfaces of the  family of the lower saddle (black line), top family (red line) and bottom family (green line), respectively. Note that we did not present the range  of $p_y$ coordinate because it is determined from the other three coordinates through the Hamiltonian of the system. In the panels A, B and C the range of the  coordinates of the top family coincide with the range of the  same coordinates  of the bottom family. Moreover the range of the $y$ coordinate of the family of the lower saddle is zero for all energies (see panel B). Finally,  the continuation of the range of the $p_x$ coordinate of the families is evident as the energy varies (see panel C).

\end{enumerate}

%2D projections of periodic orbits  
%X-Y
%(Central: E=-1.2, -0.6, -0.2, -0.00055)

%(top: E = -0.00055, 0.2, 0.4, 0.65)

%(bottom: E = -0.00055, 0.2, 0.4, 0.65)

%In Figures \ref{divcentralxy}-\ref{divbottomypy}

%2D projections of dividing surface 
%X-Y
%(Central: E=-1.2, -0.6, -0.2, -0.00055)

%(top: E = -0.00055, 0.2, 0.4, 0.65)

%(bottom: E = -0.00055, 0.2, 0.4, 0.65)

%X-PX
%(Central: E=-1.2, -0.6, -0.2, -0.00055)

%(top: E = -0.00055, 0.2, 0.4, 0.65)

%(bottom: E = -0.00055, 0.2, 0.4, 0.65)

%Y-PY
%(Central: E=-1.2, -0.6, -0.2, -0.00055)

%(top: E = -0.00055, 0.2, 0.4, 0.65)

%(bottom: E = -0.00055, 0.2, 0.4, 0.65)

\section{Conclusions}
\label{conclusions}
We studied  a bifurcation of  periodic orbit dividing surfaces that occurs because of a pitchfork bifurcation of periodic orbits in a nonlinear Hamiltonian system with two degrees of freedom. We investigated the structure, the range and the minimum and maximum of the periodic orbit dividing surfaces of the family of periodic orbits of the lower saddle of our system and of its bifurcating families of periodic orbits. Our main conclusions are:
\begin{enumerate}
    \item The dividing surfaces of the family of  periodic orbits of the lower saddle are presented as  ellipsoids and  filamentary structures in the 3D subspaces of the phase space. The ellipsoids become sharper as we increase the energy.
    
    \item The dividing surfaces of the bifurcating families are presented as ellipsoids  with a sharp edge  or as curved filamentary structures. The sharpness of the ellipsoids  decreases as we increase the energy.
    
    \item The dividing surfaces of the bifurcating families coincide with the dividing surfaces of the family of the lower saddle in the $(x,p_x,p_y)$ subspace of the phase space. 
    
    \item The dividing surfaces of the bifurcating families  are on both sides   of the dividing surfaces  of the family of the lower saddle in the 3D subspaces of the phase space $(x,y,p_x)$, $(x,y,p_y)$ and $(y,p_x,p_y)$. This happens until the disappearance of the dividing surfaces of the family of lower saddles for values of energy from 0 (for which we have the appearance of the high energy saddle)  and above. As we increase the energy, the dividing surfaces of the bifurcating families move away from each other.
    
    \item The ranges of the dividing surfaces  of the bifurcating families coincide with each other and they follow the evolution of the range of the dividing surfaces of the lower saddle (except  the range in the $y$ coordinate where the range of the dividing surfaces of the lower saddle is 0).
    
    \item  The minimum and the maximum of the dividing surfaces coincide with each other with respect to the $x$ and $p_x$ coordinate (the $p_y$ coordinate is determined through the Hamiltonian of the system) and they follow, in these cases, the evolution of the minimum and maximum of the dividing surfaces of the lower saddles.  
\end{enumerate}

\section*{Acknowledgments}

The authors acknowledge the financial support provided by the EPSRC Grant No. EP/P021123/1 and MA also acknowledges  support  from  the  grant  “Juan  de  la  Cierva  Incorporaci{\'o}n”  (IJC2019-040168-I).

\clearpage
\bibliography{main}

\end{document}